\documentstyle[12pt]{article}
\begin{document}

\tolerance=5000

\def\pp{{\, \mid \hskip -1.5mm =}}
\def\cL{{\cal L}}
\def\be{\begin{equation}}
\def\ee{\end{equation}}
\def\bea{\begin{eqnarray}}
\def\eea{\end{eqnarray}}
\def\tr{{\rm tr}\, }
\def\nn{\nonumber \\}
\def\e{{\rm e}}
\def\D{{D \hskip -3mm /\,}}

\  \hfill
\begin{minipage}{3.5cm}
March 2003 \\
\end{minipage}

\vfill

\begin{center}
{\large\bf Quantum deSitter cosmology and phantom matter}

\vfill

{\sc Shin'ichi NOJIRI}\footnote{nojiri@cc.nda.ac.jp}
and {\sc Sergei D. ODINTSOV}$^{\spadesuit}$\footnote{
odintsov@mail.tomsknet.ru Also at ICREA and IEEC, Barcelona, Spain}

\vfill

{\sl Department of Applied Physics \\
National Defence Academy,
Hashirimizu Yokosuka 239-8686, JAPAN}

\vfill

{\sl $\spadesuit$
Lab. for Fundamental Studies,
Tomsk State Pedagogical University,
634041 Tomsk, RUSSIA}

\vfill

{\bf ABSTRACT}

\end{center}

We consider deSitter universe and Nariai universe induced by 
quantum CFT with classical phantom matter and perfect fluid.
The model represents the combination of trace-anomaly driven 
inflation and phantom driven deSitter universe. 
The similarity of phantom matter with 
quantum CFT indicates that phantom scalar may be the effective 
description for some quantum field theory.
It is demonstrated that it is easier to achieve the acceleration 
of the scale factor preserving the energy conditions in such unified model.
Some properties of unified theory (anti-gravitating solutions,
negative ADM mass Nariai solution, relation with steady state)
are briefly mentioned.

\vfill


\newpage

1. Recent astrophysical data indicate to the acceleration of the scale
factor of
the observable universe \cite{SuNv}. There is number of scenarios 
(see \cite{refs} and refs. therein) where it is considered 
the dark energy, 
which generates the acceleration. 
One simple possibility to model such the accelerating 
scale factor is to introduce the (phantom) matter with negative energy density
\cite{negative}. Such phantom matter may serve as another candidate for
dark energy. However, due to number of problems (violation of energy
conditions 
and related negative energy density) it does not seem to be quite realistic 
dark energy. In this respect, several questions appear. Can one suggest 
the natural mechanism of generation of phantom matter  at the early Universe?
Is it possible to construct the cosmological model where positive aspects 
of phantom still survive and negative aspects somehow disappear? 

The strange properties of the phantom scalar 
(with negative kinetic energy) in the space with non-zero cosmological
constant have been recently discussed in the very interesting paper by Gibbons
\cite{Gibbons}. In the present letter we consider some generalization of 
phantom cosmology, i.e. the situation where classical matter contains 
perfect fluid and phantom scalar but there is also quantum contribution.
The quantum effects are described via the account of conformal anomaly,
reminding about anomaly-driven inflation \cite{starobinsky}.  
Quantum effects may lead also to negative energy density 
(for higher derivative conformal matter) or to negative pressure (usual
matter)
what may indicate that phantom corresponds to the effective description of
some QFT.
Then, it suggests some mechanism to introduce the phantom at early Universe.
On the same time, it is easier to get the consistent deSitter-like universe
(even with zero cosmological constant) due to unified theory: phantom
and QFT. 
This is again quite interesting: acceleration of the scale factor is achieved 
due to phantom, but most of energy conditions are satisfied due to quantum
effects!
We also describe the consistent Nariai universe induced by such unified
model. Again, like in pure phantom cosmology, it is shown
that quantum effects may generate the negative ADM mass solution.

2. Let us start from the Einstein equation with the phantom (scalar) field $C$
\cite{Gibbons}
\be
\label{phtm1}
R_{\mu\nu} - {1 \over 2}Rg_{\mu\nu}
=8\pi G \left\{ \left(\rho + p\right)U_\mu U_\nu + p g_{\mu\nu} 
 - \partial_\mu C \partial_\nu C + {1 \over 2}g_{\mu\nu} g^{\alpha\beta}
\partial_\alpha C \partial_\beta C\right\}\ .
\ee
Let the metric of the 4 dimensional spacetime has the warped form:
\be
\label{ddS1}
ds^2= - dt^2 + L^2\e^{2A}\sum_{i,j=1}^3\tilde g_{ij}dx^i dx^j\ .
\ee
Here $\tilde g_{ij}$ is the metric which satisfies 
$\tilde R_{ij}=k \tilde g_{ij}$ with the Ricci curvature $\tilde R_{ij}$ 
given by $\tilde g_{ij}$ and $k=0,\pm 2$.
The simplest way to account for quantum effects 
(at least, for conformal matter) is to include the contributions from the
conformal anomaly:
\be
\label{OVII}
T=b\left(F+{2 \over 3}\Box R\right) + b' G + b''\Box R\ ,
\ee
where $F$ is the square of 4d Weyl tensor, $G$ is Gauss-Bonnet invariant,
which are
given as
\bea
\label{GF}
F&=&{1 \over 3}R^2 -2 R_{ij}R^{ij}+ R_{ijkl}R^{ijkl} \nn
G&=&R^2 -4 R_{ij}R^{ij}+ R_{ijkl}R^{ijkl} \ ,
\eea
In general, with $N$ scalar, $N_{1/2}$ spinor, $N_1$ vector fields, $N_2$
($=0$ or $1$) 
gravitons and $N_{\rm HD}$ higher derivative conformal scalars, $b$, $b'$
and $b''$ are 
given by
\bea
\label{bs}
&& b={N +6N_{1/2}+12N_1 + 611 N_2 - 8N_{\rm HD} 
\over 120(4\pi)^2}\nn 
&& b'=-{N+11N_{1/2}+62N_1 + 1411 N_2 -28 N_{\rm HD} 
\over 360(4\pi)^2}\ , \quad b''=0\ .
\eea

The contributions due to conformal anomaly to $\rho$ and $p$ are found in
\cite{NOev,NOOfrw}, 
\bea
\label{hhrA3}
\rho_A&=&-\left.{1 \over a^4}\right[b'\left( 6 a^4 H^4 + 12 a^2 H^2\right) \\
&& + \left({2 \over 3}b + b''\right)\left\{ a^4
\left( -6 H H_{,tt}- 18 H^2 H_{,t} + 3 H_{,t}^2 \right) + 6 a^2 H^2\right\}
\nn
&&  -2b +6 b' -3b'' \Bigr] ,\nn
\label{hhrAA1}
p_A&=&b'\left\{ 6 H^4 + 8H^2 H_{,t} + {1 \over a^2}\left( 4H^2 + 8
H_{,t}\right) \right\} \nn
&& \left.+ \left({2 \over 3}b + b''\right)\right\{ -2H_{,ttt} -12 
H H_{,tt} - 18 H^2 H_{,t} - 9 H_{,t}^2 \nn
&& \left. + {1 \over a^2} 
\left( 2H^2 + 4H_{,t}\right) \right\} - { -2b +6 b' -3b''\over 3a^4} \ .
\eea
Here, the ``radius'' of the universe $a$ and the Hubble parameter $H$ are
\be
\label{en6}
a\equiv L\e^A\ ,\quad H={1 \over a}{d a \over dt}
={d A \over dt}\ .
\ee
One may also set 
\be
\label{ph1}
U_\mu = \delta^t_{\ \mu}\ ,\quad C=\tilde a t + \tilde b\ .
\ee
The latter is the solution of the equation of motion of the phantom field:
\be
\label{ph2}
0=\Box C = - \partial_t^2 C\ .
\ee
Note that $\tilde a$ and $\tilde b$ are arbitrary then there are infinitely many 
solutions (distributions of phantom matter) of (\ref{ph2}). 
In order to simplify the discussion, the deSitter space is taken below in
the following form (instead of (\ref{ddS1}):
\be
\label{dS1}
ds^2= - dt^2 + L^2 \cosh^2 {t \over L}d\Omega_3^2\ .
\ee
Then , calculating quantum energy density and pressure one gets
\be
\label{phtm2}
\rho_A=-p_A = - {6b' \over L^4}\ .
\ee
Since for the metric (\ref{dS1}), 
\be
\label{phtm2b}
R_{\mu\nu}= {3 \over L^2}g_{\mu\nu}\ ,\quad R={12 \over L^2}\ ,
\ee
the $(tt)$ and $(ij)$-components in the Einstein equations (\ref{phtm1}) 
with account of quantum effects are, respectively:
\bea
\label{phtm3}
{3 \over L^2}&=&8\pi G\left(\rho_{\rm matter} - {6b' \over L^4} - {\tilde a^2
\over 2}\right)\ ,\\
\label{phtm4}
 -{3 \over L^2}&=&8\pi G\left(p_{\rm matter} + {6b' \over L^4} - {\tilde a^2 \over
2}\right)\ .
\eea
Here $\rho_{\rm matter}$ and $p_{\rm matter}$ are contributions from the
matter to the energy density $\rho$ and pressure $p$.
One sees that for higher-derivative scalar, like for classical phantom, the
quantum energy 
density is negative. On the same time, for usual matter the quantum energy
is positive,
while quantum pressure is negative. Hence, in principle, phantom field may
be some effective
theory coming from more fundamental QFT. In other words, qualitatively 
we suggested the way how phantom may appear at the early Universe. 

By combining (\ref{phtm3}) and (\ref{phtm4}), we obtain
\bea
\label{phtm5}
0&=& \rho_{\rm matter} + p_{\rm matter} - \tilde a^2 \ ,\\
\label{phtm6}
0&=& {12b' \over L^4} + {6 \over 8\pi G L^2} - \rho_{\rm matter} + p_{\rm
matter} \ ,
\eea
which tells that there is a dS solution even if there is no cosmological
constant. Eq.(\ref{phtm6}) has solutions 
\be
\label{phtm6b}
{1 \over L^2}={1 \over 12b'}\left\{ - {3 \over 4\pi G}\pm 
\sqrt{\left({3 \over 8\pi G}\right)^2 + 12b'\left(\rho_{\rm matter} -
p_{\rm matter}\right)}
\right\}\ ,
\ee
with respect to $L^2$ if
\be
\label{phtm7}
\left({3 \over 8\pi G}\right)^2 + 12b'\left(\rho_{\rm matter} - p_{\rm
matter}\right)\geq 0\ .
\ee 
We should note the $+$-sign in $\pm$ in (\ref{phtm6b}) corresponds to the
solution of ref.\cite{Gibbons} in the limit $b'\to 0$. On the other hand,
when 
$\rho_{\rm matter} - p_{\rm matter}=0$, the solution with $-$-sign in $\pm$
in (\ref{phtm6b}) corresponds to the anomaly driven inflation by
Starobinsky\cite{starobinsky}. 
The limiting case $\rho_{\rm matter} = p_{\rm matter}$ describes the 
stiff matter. Without quantum effects, there is no non-trivial solution for 
$L^2$ \cite{Gibbons} but due to the conformal anomaly, there is a
nontrivial solution even for the stiff matter.  Having now the explicit
deSitter 
cosmological solution, one can address the question: how the energy
conditions look in our model?

There are several standard types of the energy conditions in cosmology:
\begin{itemize}
\item Null Energy Condition (NEC):
\be
\label{phtm11}
\rho + p \geq 0
\ee
\item Weak Energy Condition (WEC):
\be
\label{phtm8}
\rho\geq 0 \ \mbox{and}\ \rho + p \geq 0 
\ee
\item Strong Energy Condition (SEC):
\be
\label{phtm9}
\rho + 3 p \geq 0\ \mbox{and}\ \rho + p \geq 0
\ee
\item Dominant Energy Condition (DEC):
\be
\label{phtm10}
\rho\geq 0 \ \mbox{and}\ \rho \pm p \geq 0 
\ee
\end{itemize}
For the present case, from Eqs.(\ref{phtm3}) and (\ref{phtm4}), we have
\bea
\label{phtm12}
\rho_{\rm matter} &=& {6b' \over L^4} + {\tilde a^2 \over 2} + {3 \over 8\pi G
L^2} ,\\
\label{phtm13}
p_{\rm matter} &=& -{6b' \over L^4} + {\tilde a^2 \over 2} - {3 \over 8\pi G L^2} \ .
\eea
NEC is always satisfied from (\ref{phtm5}):
\be
\label{phtm14}
\rho_{\rm matter} + p_{\rm matter}\geq 0\ .
\ee
WEC could be satisfied, from (\ref{phtm12}), if 
\bea
\label{phtm15}
&& {6b' \over L^4} + {\tilde a^2 \over 2} + {3 \over 8\pi G L^2} \nn
&=& {\tilde a^2 \over 2L^4}\left\{ L^2 - {1 \over \tilde a^2}\left( - {3 \over 8\pi G} + 
\sqrt{\left({3 \over 8\pi G}\right)^2 - 12b'}\right)\right\} \nn
&& \times \left\{ L^2 - {1 \over \tilde a^2}\left( - {3 \over 8\pi G} - 
\sqrt{\left({3 \over 8\pi G}\right)^2 - 12b'}\right)\right\} \nn
&\geq& 0\ .
\eea
If $b'<0$ as usually, the quantity inside the square root is always positive. 
Eq.(\ref{phtm15}) gives a non-trivial constraint for the length parameter
$L^2$
\be
\label{phtm16}
L^2 \geq {1 \over \tilde a^2}\left( - {3 \over 8\pi G} + 
\sqrt{\left({3 \over 8\pi G}\right)^2 - 12b'}\right)\ .
\ee
In case of no quantum effects ($b'=0$), the constraint becomes trivial:
$L^2\geq 0$. Since 
\bea
\label{phtm17}
\rho_{\rm matter} + 3p_{\rm matter} &=& -{12b' \over L^4} + 2\tilde a^2 - {3 \over
4\pi G L^2} \nn 
&=& {2\tilde a^2 \over2L^4}\left\{ L^2 - {1 \over \tilde a^2}\left( {3 \over 16\pi G} + 
\sqrt{\left({3 \over 16\pi G}\right)^2 + 6b'}\right)\right\} \nn
&& \times \left\{ L^2 - {1 \over \tilde a^2}\left( {3 \over 16\pi G} - 
\sqrt{\left({3 \over 16\pi G}\right)^2 + 6b'}\right)\right\} \ ,
\eea
if 
\be
\label{phtm18}
\left({3 \over 16\pi G}\right)^2 + 6b'\leq 0\ ,
\ee
the quantity inside the square root in (\ref{phtm17}) is non-positive and
we find the SEC 
is always satisfied; $\rho_{\rm matter} + 3p_{\rm matter}\geq 0$. On the
other hand, if  
\be
\label{phtm19}
\left({3 \over 16\pi G}\right)^2 + 6b'> 0\ ,
\ee
the SEC gives the non-trivial constraint for $L^2$
\bea
\label{phtm20}
&& 0\leq L^2 \leq {1 \over \tilde a^2}\left( {3 \over 16\pi G} - 
\sqrt{\left({3 \over 16\pi G}\right)^2 + 6b'}\right)\ \nn
\mbox{or}\ && L^2 \geq {1 \over \tilde a^2}\left( {3 \over 16\pi G} + 
\sqrt{\left({3 \over 16\pi G}\right)^2 + 6b'}\right)\ .
\eea
The above contraint becomes trivial ($L^2\geq 0$) again if we do not
include the conformal 
anomaly. Since $b'$ is negative for the usual matter fields,
Eq.(\ref{phtm18}) tells
that if the quantum effect is large, the SEC could be satisfied more
easily. On the other 
hand if the quantum effect is small but does not vanish, the SEC may not be
satisfied. 

Eq.(\ref{phtm7}) can be rewritten, if $b'<0$,  as 
\be
\label{phtm21}
\rho_{\rm matter} - p_{\rm matter} \leq - { 1\over 12b'}\left({3 \over 8\pi
G}\right)^2 \ ,
\ee
which does not conflict with the DEC but (\ref{phtm21}) gives non-trivial
constraint 
for the matter field. This constraint does not appear if we do not include
the 
contribution from the conformal anomaly. Then due to the quantum effect,
there might happen that the DEC could not be satisfied. 

The contributions to $\rho$ and $p$ from the phantom field $C$ is, from 
(\ref{phtm3}) and (\ref{phtm4}), given by
\be
\label{phtm23}
\rho_C=p_C = -{\tilde a^2 \over 2}\ .
\ee
Therefore any energy conditions cannot be satisfied for pure phantom unless
$\tilde a=0$. When $\tilde a=0$, from (\ref{phtm5}) it follows
\be
\label{phtm24}
0= \rho_{\rm matter} + p_{\rm matter} \ ,
\ee
which is limiting case but does not violate any energy condition although
this would require negative pressure. Thus, the energy conditions when
quantum CFT presents are satisfied, unlike to the case of pure phantom.

The condition that the universe accelerates is that $L^2$ (\ref{phtm6b}) is
real. Therefore 
the condition is given by (\ref{phtm7}), which can be rewritten as
(\ref{phtm21}) if $b'<0$ 
as usual. If $b'>0$, instead of (\ref{phtm21}), one gets 
\be
\label{phtm22}
\rho_{\rm matter} - p_{\rm matter} \geq - { 1\over 12b'}\left({3 \over 8\pi
G}\right)^2 \ .
\ee
We may rewrite $(tt)$ component of Eq.(\ref{phtm1}) in the form of the FRW
equation: 
\be
\label{FRW1}
H^2={8\pi G \over 3}\rho - {k \over 2 a^2} \ .
\ee
Here the energy density $\rho$ is a sum of the contributions from the
matter, phantom field, 
and the conformal anomaly. The energy density $\rho_C$ (\ref{phtm23}) from
the phantom field 
is always constant. On the other hand, the energy density $\rho_A$
(\ref{hhrA3}) from the
conformal anomaly contains the higher power of the Hubble parameter $H$ and
its higher 
derivatives. Although $\rho_A$ is a constant in case of deSitter space, if
the universe 
expands very rapidly, $\rho_A$ mainly contributes to the expansion.   

Eq.(\ref{phtm1}) tells that the total energy momentum tensor (EMT) 
$T_{\mu\nu} = T_{{\rm matter}\,\mu\nu} + T_{C\,\mu\nu} + T_{A\,\mu\nu}$ is
conserved 
$\nabla^\mu T_{\mu\nu}=0$ due to the Bianchi identiy. As in \cite{Gibbons},
each term of 
the EMT is not conserved. In \cite{Gibbons}, if the EMT $T_{C\,\mu\nu}$
coming from the 
phantom field $C$ varies non-trivially, the steady state observer will
observe the 
creation of the matter. The EMT $T_{A\,\mu\nu}$ defined by the conformal
anomaly has 
been constructed in (\ref{hhrA3}) to be conserved. Then the  particle
creation can occur if
the phantom field becomes non-trivial. Since the energy density $\rho_C$ in
(\ref{phtm23}) 
from the phantom field, Eq.(\ref{phtm12}) tells that the matter energy
density 
$\rho_{\rm matter}$ is also constant although the universe is expanding.
This implies 
the steady creation of the matter. Taking into account the similarity of
phantom
with quantum matter, it may not look so strange. It is known that
quantum particles creation occurs in the early universe.  

In \cite{Gibbons}, the arguments are given that the phantom field $C$
generates repulsive 
gravitational force (anti-gravity) and there appears multi-objects static
solution as 
\bea
\label{anti1}
&& ds^2 = - \e^{U({\bf x})} dt^2 + \e^{-2U({\bf x})}d{\bf x}^2\ ,\nn
&& \kappa \left(C - C_\infty\right)=U=-\sum_i{M_i \over \left|{\bf x} -
{\bf x}_i\right|}\ .
\eea
Here $\kappa^2=4\pi G$ and $M_i$'s are constants, which can be negative or
positive. Even 
with quantum effects, such a solution could exist when we can regard that
the quantum 
effects are small. In an exotic case that $b'>0$, the energy density
(\ref{phtm2}) coming 
from the quantum effect becomes negative although the pressure is positive.
Then the 
quantum effect itself may become a source of the anti-gravity. As discussed
in 
\cite{Gibbons}, such a negative mass particle corresponding to the negative
ADM mass 
solution chases the positive mass object. Then one may find such a exotic
particle 
arround the black hole. Such a negative mass solution can be found exactly
without the 
quantum correction. If the quantum correction is small and can be treated
perturbatively, 
the qualitative structure would not be changed and such a solution could
exist even if we 
include the quantum correction. 

3. Another interesting example is the Nariai space, which is the extreme
limit of the 
Schwarzschil-deSitter black hole, whose metric is given by the sum of 2d
deSitter (dS$_2$) 
and 2d sphere S$^2$:
\be
\label{Nr1}
ds^2 = L^2\left(ds_{{\rm dS}_2}^2 + d_{{\rm S}^2}^2\right)\ .
\ee
Then 
\be
\label{Nr2}
F={16 \over 3L^4}\ ,\quad G={8 \over L^4}\ ,
\ee
and therefore the conformal anomaly is given by
\be
\label{Nr3}
T={1 \over L^4}\left({4 \over 3}b + 2b'\right)\ .
\ee
Approximately, the quantum EMT caused by the conformal anomaly (forgetting
about vacuum 
dependent piece) is given by
\be
\label{Nr4}
T_{\mu\nu}^A = {T \over 4}g_{\mu\nu}={1 \over L^4}\left({b \over 3} +
{b'\over 2}\right)
g_{\mu\nu}\ .
\ee
When we choose the metric of 2d deSitter as
\be
\label{Nr5}
L^2 ds_{{\rm dS}_2}^2 = -dt^2 + \e^{2t \over L}dx^2\ ,
\ee
we may assume (\ref{ph1}) again. Then  the $(tt)$ and $(ij)$ components of 
(\ref{phtm1}) are given by
\bea
\label{Nr6}
 - {1 \over L^2}&=&8\pi G \left\{ -\rho_{\rm matter} + {\tilde a^2 \over 2} 
+ {1 \over L^4}\left({b \over 3} + {b' \over 2}\right)\right\} \ ,\nn
 - {1 \over L^2}&=&8\pi G \left\{ p_{\rm matter} - {\tilde a^2 \over 2} 
+ {1 \over L^4}\left({b \over 3} + {b' \over 2}\right)\right\} \ .
\eea
If we analytically continue $L^2$ by $-L^2$, the spacetime metric
(\ref{Nr1}) is replaced by the sum of 2d anti-deSitter space and 2d
hyperboloid: 
\be
\label{Nr7}
ds^2 = L^2\left(ds_{{\rm AdS}_2}^2 + d_{{\rm H}^2}^2\right)\ ,
\ee
which can be obtained from Schwarzschild-AdS black hole with negative ADM
mass by the Nariai-like limit. With the metric of 2d anti-deSitter as
\be
\label{Nr8}
L^2 ds_{{\rm AdS}_2}^2 = dx^2 - \e^{2x \over L}dt^2\ ,
\ee
Eq.(\ref{ph1}) is a solution again. Then instead of (\ref{Nr6}) one gets  
\bea
\label{Nr10}
{1 \over L^2}&=&8\pi G \left\{ -\rho_{\rm matter} + {\tilde a^2 \over 2} 
+ {1 \over L^4}\left({b \over 3} + {b' \over 2}\right)\right\} \ ,\nn
{1 \over L^2}&=&8\pi G \left\{ p_{\rm matter} - {\tilde a^2 \over 2} 
+ {1 \over L^4}\left({b \over 3} + {b' \over 2}\right)\right\} \ .
\eea
As in (\ref{phtm5}) and (\ref{phtm6}),  (\ref{Nr10}) may be written as 
\bea
\label{Nr11}
0&=&\rho_{\rm matter} + p_{\rm matter} - \tilde a^2 \ ,\\
\label{Nr12}
{2 \over L^2}&=&8\pi G\left\{- \rho_{\rm matter} + p_{\rm matter} 
+ {2 \over L^4}\left({b \over 3} + {b' \over 2}\right)\right\} \ .
\eea
We should note that when $b'> -{2 \over 3}b$, which is a little bit exotic,
even if 
$\rho_{\rm matter}=p_{\rm matter}=\tilde a=0$, there is a nontrivial solution of
(\ref{Nr12}):
\be
\label{Nr13}
L^2 = - 8\pi G \left({b \over 3} + {b' \over 2}\right)\ , 
\ee
which indicates that the quantum effects could generate the negative ADM
mass solution. It follows  that the unified model (with perfect fluid,
phantom and quantum CFT) gives the possibility of the creation of Nariai
universe.  

Thus, we demonstrated that phantom scalar in many respects looks like
strange effective 
quantum field theory. Moreover, when matter is composed of phantom, perfect
fluid and 
quantum CFT it is somehow easier to realize the accelarating deSitter-like
universe, 
while most of energy conditions may be preserved. It would be interesting to 
estimate the cosmological applications of phantom field non-minimally
coupled with gravity.

\ 

\noindent
{\bf Acknowledgments}

The work by S.N. is supported in part by the Ministry of Education, 
Science, Sports and Culture of Japan under the grant n. 13135208.

\end{document}